\documentclass[journal]{IEEEtran}

\usepackage{cite}
\usepackage{amsmath,amssymb,amsfonts}
\usepackage{algorithmic}
\usepackage{graphicx}
\usepackage{textcomp}
\usepackage{xcolor}
\usepackage{cuted}
\usepackage{stfloats}
\usepackage{lineno,hyperref}
\usepackage{color,xcolor,colortbl}
\usepackage{multirow}
\usepackage[caption=false,font=footnotesize]{subfig}
\usepackage[linesnumbered,ruled,vlined]{algorithm2e}
\SetKwInput{KwInput}{Input}
\SetKwInput{KwOutput}{Output}
\usepackage{bigstrut}
\usepackage{array}
\usepackage[font=small]{caption}

\def\BibTeX{{\rm B\kern-.05em{\sc i\kern-.025em b}\kern-.08em
    T\kern-.1667em\lower.7ex\hbox{E}\kern-.125emX}}

\begin{document}

\title{Robustness and Security Enhancement of Radio Frequency Fingerprint Identification in Time-Varying Channels}

\author{Lu~Yang,~\IEEEmembership{Student~Member,~IEEE,}
        Seyit~Camtepe,~\IEEEmembership{Senior~Member,~IEEE,}
        Yansong~Gao,~\IEEEmembership{Senior~Member,~IEEE,}
        Vicky~Liu, and~Dhammika~Jayalath,~\IEEEmembership{Senior~Member,~IEEE}
\thanks{This article was presented in part at the IEEE PIMRC 2023~\cite{yang2023use}.}
\thanks{\textit{(Corresponding author: Lu Yang)}}
\thanks{Lu~Yang and Dhammika~Jayalath are with the School of Electrical Engineering and Robotics, Queensland University of Technology, Brisbane, Australia (email: L41.yang@hdr.qut.edu.au; dhammika.jayalath@qut.edu.au).}
\thanks{Seyit~Camtepe and Yansong~Gao are with Data61, CSIRO, Sydney, Australia (email: seyit.camtepe@data61.csiro.au; garrison.gao@data61.csiro.au).}
\thanks{Vicky~Liu is with School of Computer Science, Queensland University of Technology, Brisbane, Australia (email: v.liu@qut.edu.au).}
\thanks{Manuscript received XXX XX, 2024; revised XXX XX, 2024.}}

\maketitle

\begin{abstract}
Radio frequency fingerprint identification (RFFI) is becoming increasingly popular, especially in applications with constrained power, such as the Internet of Things (IoT). Due to subtle manufacturing variations, wireless devices have unique radio frequency fingerprints (RFFs). These RFFs can be used with pattern recognition algorithms to classify wireless devices. However, Implementing reliable RFFI in time-varying channels is challenging because RFFs are often distorted by channel effects, reducing the classification accuracy. This paper introduces a new channel-robust RFF, and leverages transfer learning to enhance RFFI in the time-varying channels. Experimental results show that the proposed RFFI system achieved an average classification accuracy improvement of 33.3\% in indoor environments and 34.5\% in outdoor environments. This paper also analyzes the security of the proposed RFFI system to address the security flaw in formalized impersonation attacks. Since RFF collection is being carried out in uncontrolled deployment environments, RFFI systems can be targeted with false RFFs sent by rogue devices. The resulting classifiers may classify the rogue devices as legitimate, effectively replacing their true identities. To defend against impersonation attacks, a novel keyless countermeasure is proposed, which exploits the intrinsic output of the softmax function after classifier training without sacrificing the lightweight nature of RFFI. Experimental results demonstrate an average increase of 0.3 in the area under the receiver operating characteristic curve (AUC), with a 40.0\% improvement in attack detection rate in indoor and outdoor environments.
\end{abstract}


\section{Introduction}\label{sec:introduction}

\IEEEPARstart{T}{he} rapid increase of the Internet of Things (IoT) devices has created an urgent need for better security measures in critical IoT applications~\cite{hassija2019survey}. One important aspect of securing devices involves authenticating them, which entails identifying unauthorized devices and categorizing registered ones~\cite{yang2017survey}. Due to high computational costs, traditionally used public key cryptography (PKC) algorithms are cumbersome for securing IoT devices. Further, PKC requires a certification authority for key sharing, which may only sometimes be available due to the large number and wide distribution of IoT devices~\cite{xu2015device}.

A lightweight approach for authenticating IoT devices is demanded. One such method is radio frequency fingerprint identification (RFFI), which has garnered significant research attention~\cite{riyaz2018deep,zhang2019physical,tian2019new,sankhe2020no,rajendran2020injecting,xie2021generalizable,ng2022compressive,soltanieh2020review,yang2023led}. RFFI exploits unique hardware characteristics that result from manufacturing imperfections, causing slight distortions in transmitted signals. Like biometric characteristics recognize humans, these subtle features are distinctive and hard to replicate physically. Thus, receivers can extract and verify these features against pre-shared feature information for device authentication. Since RFFI generates little overhead to end devices with no computationally intensive algorithms implemented, it is energy-efficient and desirable for power-constrained device authentication.

An RFFI classifier is a model trained using radio frequency fingerprints (RFFs). Deep learning is widely implemented, which is beneficial in automatically identifying and extracting RFFs from received signals~\cite{merchant2018deep,das2018deep,peng2020deep,roy2020rfal,he2020cooperative,peng2022radio,liu2023radio}. For the network architecture, a convolutional neural network (CNN) is widely used for pattern recognition tasks, making it well-suited for fingerprint identification~\cite{ding2018specific,sankhe2019oracle,jian2022radio,qian2021specific,shen2021radioIFOCOM,li2022radionet,liu2022radio,xu2023polarization}. Various features are studied for authentication, such as in-phase and quadrature samples~\cite{sun2020radio,zhou2021robust,zeng2023multi}, Fourier transform results~\cite{robyns2017physical,guo2021specific,xing2023design}, and spectrogram~\cite{shen2021radioJEAC,reising2021radio}. The spectrogram-based model has demonstrated the highest device classification accuracy, motivating us to adopt spectrogram and deep learning classifiers to benchmark the research.

RFFs obtained directly from received signals are susceptible to wireless channel effects, such as path loss, multipath fading, and the Doppler effect. These effects significantly diminish the accuracy of device authentication in RFFI systems~\cite{pourkabirian2022robust,yang2022channel,fadul2021nelder,soltani2020rf,al2020exposing,restuccia2019deepradioid,qi2023lightweight,he2023radio}. We overcome this by using the power amplifier (PA) nonlinearity quotient to mitigate channel effects as an environmentally robust feature. This feature is calculated by applying the convolution theorem in the frequency domain. In environments with fast-moving objects, multipath fading and the Doppler effect are more pronounced, especially for low data rate power-constrained devices. Researchers used data augmentation to train classifiers with simulated datasets featuring multipath fading and the Doppler effect to address the issue~\cite{shen2022towards,wang2022radio,al2021deeplora,soltani2020more,shen2023towards}. However, these simulated datasets lack knowledge of the real deployment environments and substantially increase the disk and memory storage requirements for training. We circumvent this by leveraging transfer learning to address the impact of fast-moving objects. Unlike simulated data augmentation, transfer learning requires much less data for classifier training, reducing required disk memory and storage. It is developed by training a base model with undistorted RFFs, followed by the retraining with distorted RFFs from deployment environments. This enables the model to acknowledge original RFFs and channel effects introduced by fast-moving objects~\cite{sharaf2016authentication,wang2020radio,kuzdeba2021transfer}.

We conducted extensive indoor and outdoor experiments investigating the PA nonlinearity quotient and transfer learning classifier. The results show that the proposed classifier outperformed conventional spectrogram and deep learning classifiers.

The paper also aims to examine the security of RFFI in uncontrolled deployment environments. We formalized two impersonation attacks that target RFFI systems in enrollment and authentication steps. Our findings indicate that RFFI systems are susceptible to impersonation attacks during enrollment. To address this vulnerability without compromising the lightweight nature of RFFI, we proposed a new keyless technique that leverages the intrinsic output from the softmax function in classifier training for attack detection. We conducted extensive experiments to validate the proposed countermeasure. The results confirm that the countermeasure can detect the impersonation attack, thereby enhancing the security of RFFI systems in uncontrolled deployment environments. In summary, our contributions are threefold.

\begin{itemize}
    \item We developed a new channel-robust RFF and utilized transfer learning for the first time in LoRa-based RFFI. Through extensive experiments, we validated the robustness of developing RFFI systems for accurately classifying devices in changing environments.
    \item The channel-robust RFF collection carried out in deployment environments is subjected to false RFFs from rogue devices. Therefore, we formalized the threat of impersonation attacks and investigated their impact on RFFI systems.
    \item We proposed a novel keyless countermeasure without sacrificing the lightweight nature of RFFI to address the falsing RFF challenge. The countermeasure was validated through extensive experiments to detect rogue devices effectively. Thus significantly enhancing the security of RFFI systems implemented in uncontrolled deployment environments.
\end{itemize}

The paper is structured as follows. Section~\ref{sec:PANQ} introduces the RFFI robustness enhancement. Section~\ref{sec:securityanalysis} introduces the RFFI security enhancement. Section~\ref{sec:experiments} describes the experiment setups. Section~\ref{sec:classificationanalysis} presents the experimental results of the RFFI robustness enhancement. Section~\ref{sec:impersonationattackanalysis} presents the security analysis results to enhance RFFI security. Section~\ref{sec:attackcountermeasure} introduces the countermeasure to address the RFFI vulnerability revealed in the security analysis. Section~\ref{sec:conclusion} concludes the paper.

\section{Robustness Enhancement of RFFI}\label{sec:PANQ}
\subsection{PA Nonlinearity Quotient}
Power amplifiers amplify weak signals and are an essential component of any wireless device. Several models have been proposed to accurately represent the amplitude/amplitude (AM/AM) and amplitude/phase (AM/PM) functions that describe the inherent nonlinearity of low-power and narrowband systems like IoT devices~\cite{zhu2013challenges}. The incorporation of PA nonlinearity into RFFI has been extensively explored in literature~\cite{gong2020unsupervised,polak2011identifying,zhang2016specific,hanna2019deep,satija2019specific,li2022radio,sun2022radio,he2024channel,li2024puf}. However, the practical application tends to be confined to static or semi-static channels since the performance drops significantly when wireless channels vary. Therefore, we identify the PA nonlinearity quotient, an environmentally robust feature for designing resilient RFFI systems in time-varying channels. The formalization of the PA nonlinearity quotient is as follows.

The signal in a narrowband system reaching a receiver can be given as
\begin{align*}
s(t)=h(\tau,t)\ast f_p\left [ x(t) \right ]+n(t),
\label{eq:signalModel}
\tag{1}
\end{align*}
where $x(t)$ is a baseband signal, $h(\tau,t)$ is channel impulse response, $p$ is transmission power, $f_p[\cdot]$ denotes the nonlinear effect at the transmission power, $n(t)$ is additive white Gaussian noise (AWGN), and ``$\ast$'' denotes convolution operation.

When calculating the PA nonlinearity quotient, two consecutive signals are received: one with high power and one with low power. In practical applications, IoT technologies like LoRa require packet fragmentation to increase throughput and minimize data loss caused by collisions~\cite{bor2016lora,ferre2017collision,suciu2018analysis,peper2021high}. The payload significantly limits the maximum number of LoRa sensors that can communicate on the same channel~\cite{yousuf2018throughput,lavric2019lora}. As a result, transmitting multiple consecutive signals with fragments is practical in LoRa networks. Further, it should be noted that the power alternation does not impact the LoRa data rate and can keep a minimal impact on the normal communications~\cite{bor2017lora}. In practice, fixed LoRa transmission power is set to the maximum level by default, and novel techniques have been proposed to dynamically change the power level to increase energy efficiency and outperform the adaptive data rate (ADR) algorithms in the legacy LoRaWAN protocol~\cite{li2020dylora,al2021optimizing}. Therefore, carefully designed power alternation that involves low transmission power can generate the PA nonlinearity quotient and extend the device's operation time simultaneously. The purpose of the paper is to investigate the performance of the PA nonlinearity quotient to enhance RFFI robustness. An exploration of the power alternation algorithms is excluded at this stage.

The received signals are then divided element-wise in the frequency domain. The transformation of the signals into the frequency domain is achieved with the short-time Fourier transform (STFT). The receiver's nonlinear effect is ignored since the same effect applies to all received signals. Therefore, the STFT result can be given as
\begin{flalign}
\boldsymbol{S}_p=
\begin{bmatrix}
S^{1,1}_p & S^{1,2}_p & \cdots & S^{1,M}_p \\
S^{2,1}_p & S^{2,2}_p & \cdots & S^{2,M}_p \\
\vdots & \vdots & \ddots & \vdots \\
S^{W,1}_p & S^{W,2}_p & \cdots & S^{W,M}_p
\end{bmatrix},
\label{eq:receivedspectrogram}
\tag{2}
\end{flalign}
where $p=\{h,l\}$, which denotes high power and low power. The matrix elements are given as
\begin{align*}
S_{p}^{w,m}=\sum_{n=0}^{W-1}s_{p}\left [ n \right ]g\left [ n-mR \right ]e^{-j2\pi \frac{w}{W}n}\\
\text{for }w=1,2,...,W \text{ and }m=1,2,...,M,
\label{eq:STFT}
\tag{3}
\end{align*}
where $s_{p}\left [ n \right ]$ is the discrete signal sampled by the receiver, $g\left [ n \right ]$ is the window function with length $W$, and $R$ is hop size. We implemented LoRa devices in experiments, hence $M$ is given by the LoRa configurations as
\begin{align*}
M=\frac{K\cdot \frac{2^{SF}}{B}\cdot f_S-W}{R}+1,
\label{eq:Mvalue}
\tag{4}
\end{align*}
where $K$ is the number of LoRa symbols, $SF$ is LoRa spreading factor, $B$ is LoRa bandwidth, and $f_S$ is sampling frequency of the receiver. We implemented default Hamming window settings with $W$ equal to 1024 and $R$ equal to 512 from Matlab in the experiments, and $M$ was calculated to be 319. Detailed LoRa configurations are discussed in Section~\ref{sec:experiments}.

The STFT result is further expressed as~\eqref{eq:highpowerspectrogram} for the high-power signal, where $X$ denotes the ideal spectrum of the signal, $H$ denotes channel frequency response, and $F(\cdot)$ denotes the nonlinear effect in the frequency domain. The PA nonlinearity quotient is generated using payload-independent signal fields, i.e., $X^{w,m}=X^{w,M+m}$. Therefore, the STFT result of the subsequent low-power signal is given as~\eqref{eq:lowpowerspectrogram}.

\begin{figure*}[b]
\begin{flalign}
\boldsymbol{S}_h=
\begin{bmatrix}
H^{1,1}F_{h}(X^{1,1}) & H^{1,2}F_{h}(X^{1,2}) & \cdots & H^{1,M}F_{h}(X^{1,M}) \\
H^{2,1}F_{h}(X^{2,1}) & H^{2,2}F_{h}(X^{2,2}) & \cdots & H^{2,M}F_{h}(X^{2,M}) \\
\vdots & \vdots & \ddots & \vdots \\
H^{W,1}F_{h}(X^{W,1}) & H^{W,2}F_{h}(X^{W,2}) & \cdots & H^{W,M}F_{h}(X^{W,M}) 
\end{bmatrix},
\label{eq:highpowerspectrogram}
\tag{5.1}
\end{flalign}
\end{figure*}

\begin{figure*}[b]
\begin{flalign}
\boldsymbol{S}_l=
\begin{bmatrix}
H^{1,M+1}F_{l}(X^{1,1}) & H^{1,M+2}F_{l}(X^{1,2}) & \cdots & H^{1,2M}F_{l}(X^{1,M}) \\
H^{2,M+1}F_{l}(X^{2,1}) & H^{2,M+2}F_{l}(X^{2,2}) & \cdots & H^{2,2M}F_{l}(X^{2,M}) \\
\vdots & \vdots & \ddots & \vdots \\
H^{W,M+1}F_{l}(X^{W,1}) & H^{W,M+2}F_{l}(X^{W,2}) & \cdots & H^{W,2M}F_{l}(X^{W,M}) 
\end{bmatrix}.
\label{eq:lowpowerspectrogram}
\tag{5.2}
\end{flalign}
\end{figure*}

By removing distorted samples caused by the severe Doppler effect, we assume slow and moderate fading mostly dominates wireless channels. Therefore, the channel frequency response does not change significantly during the signal duration, i.e., $H^{w,m}\approx H^{w,M+m}$. The result of an element-wise division of the received signals on the frequency domain ($\boldsymbol{Q}$) can be expressed as
\begin{flalign}
\boldsymbol{Q}=\boldsymbol{S}_h ./ \boldsymbol{S}_l=
\begin{bmatrix}
\frac{F_{h}(\boldsymbol{X^{1}})}{F_{l}(\boldsymbol{X^{1}})} & \frac{F_{h}(\boldsymbol{X^{2}})}{F_{l}(\boldsymbol{X^{2}})} & \boldsymbol\cdots & \frac{F_{h}(\boldsymbol{X^{M}})}{F_{l}(\boldsymbol{X^{M}})} \\
\end{bmatrix},
\label{eq:PAnonlinearityquotient}
\tag{6}
\end{flalign}
where ``$./$'' denotes the element-wise division and $\boldsymbol{X^m}=[X^{1,m} \quad X^{2,m} \quad \cdots \quad X^{W,m}]^T$. No channel frequency response $(H)$ presents in $\boldsymbol{Q}$. Environmentally robust RFFI is developed with $\boldsymbol{Q}$ in dB scale as,
\begin{align*}
\widetilde{\boldsymbol{Q}}=10\log_{10}(|\boldsymbol{Q}|^2).
\label{eq:QdB}
\tag{7}
\end{align*}

\subsection{Preprocessing and Generation of PA Nonlinearity Quotient }\label{sec:dataprocessing}
The PA nonlinearity quotient is generated using preambles in wireless communication packets. Preambles are used to synchronize receivers with transmitters and are widely available in different wireless technologies. Preambles generally contain a fixed number of symbols and are payload-independent, making them an ideal source for RFFI~\cite{reus2021classifying}.

We propose Algorithm~\ref{algorithm:generation} to remove severely distorted preambles caused by antenna movement and fast-moving objects, and to generate the PA nonlinearity quotient to enhance RFFI. The correlation between consecutive high-power and low-power preambles remains constant when wireless channels have no variations, since PA nonlinearity is a single-variable function impacted only by input power~\cite{zhang2021radio}. We identify the distorted preambles by detecting significantly changed correlations in practical environments compared to correlations in environments without variations, such as an anechoic chamber. After removing the distortion, an element-wise division in the frequency domain is performed to generate the PA nonlinearity quotient.

\begin{algorithm}[t]
\footnotesize
\DontPrintSemicolon
  \KwInput{$\boldsymbol{S}_{h,k},\boldsymbol{S}_{l,k}$ \quad $\%$STFT results of indoor, outdoor preambles $(h=high\:power, l=low\:power, k=indoor,\:outdoor)$}
  \KwInput{$\boldsymbol{S}_{h,c},\boldsymbol{S}_{l,c}$ \quad $\%$STFT results of anechoic chamber preambles}
  \KwInput{$\theta$ \quad $\%$ Tolerance}
  \KwOutput{$\widetilde{\boldsymbol{Q}}$ \quad $\%$ PA nonlinearity quotient (dB)}
  $\rho_k=corr\{max(\boldsymbol{S}_{h,k}),max(\boldsymbol{S}_{l,k})\}$  \quad $\%$ Correlations\\
  $\rho_c=corr\{max(\boldsymbol{S}_{h,c}),max(\boldsymbol{S}_{l,c})\}$\\
  $\rho_d=\|\rho_c-\rho_k\|$\\
  \eIf{$\rho_d\leq\theta$}
  {$\left. \boldsymbol{Q}=\boldsymbol{S}_{h,k} .\middle/ \boldsymbol{S}_{l,k} \right.$  \quad $\%$ Element-wise division\\
  $\widetilde{\boldsymbol{Q}}=10\log_{10}(|\boldsymbol{Q}|^2)$}
  {$remove \: \boldsymbol{S}_{h,k},\boldsymbol{S}_{l,k}$}
\caption{PA Nonlinearity Quotient Generation.}
\label{algorithm:generation}
\end{algorithm}

\subsection{Transfer Learning}
Transfer learning is used with the PA nonlinearity quotient to improve RFFI robustness by accounting for channel effects in real-world environments. The process of developing transfer learning-based RFFI involves the following steps.
\begin{enumerate}
\item Base model training: Train a base model using data from a large sample collected in an environment without channel effects. In our experiments, data collection is done in an anechoic chamber to test transfer learning performance. In practical scenarios, data collection can be done during device manufacturing to reduce device operation time impact.
\item Enrollment: This step is performed when devices join a network and fine-tunes the base model using data from a small sample collected in deployment environments through transfer learning. This trains a device classifier that accommodates channel effects present in practical environment samples.
\item Authentication: The device classifier is implemented to carry out RFFI in the deployment environments.
\end{enumerate}

The transfer learning-based classifier is trained with the architecture outlined in Table~\ref{table:ClassifierArchitecture}. The architecture consists of three convolution layers with 8, 16, and 32 $3\times3$ filters. A batch normalization layer and the rectified linear unit (ReLU) activation follow each convolution layer. After the activation, a $2\times2$ max pooling layer with a stride of 2 is applied. The output of the last ReLU activation is then passed to a fully connected layer. An output layer with the softmax function is placed last to generate posterior probabilities. The Adam optimizer is used to minimize losses, with a mini-batch size of 32 and an initial training rate of 0.005 that remains constant.

All $\widetilde{\boldsymbol{Q}}$ is resized to $256\times256$ with 8-bit depth as the input to enable transfer learning. The fully connected and output layers are replaced with new layers for transfer learning. The training rate is 0.0001, and the new layers' learning rate factor is configured to 20 for fine-tuning. It is important to note that the classifier architecture is straightforward and can be further adjusted to enhance RFFI robustness. However, the primary goal of this work is to demonstrate the effectiveness of transfer learning rather than optimizing the classifier.

\begin{table}[t]
\footnotesize
\centering
\caption{Classifier Layers, Parameters, and Activation}
\label{table:ClassifierArchitecture}
\begin{tabular}{|c|c|c|c|}
\hline
Layer           & Dimension             & Parameters    & Activation\\\hline
Input           & $256\times256$        & ---           & ---\\       \hline
Convolution, BN & $8\times(3\times3)$   & 80, 16        & ReLU\\      \hline
MaxPooling      & $2\times2$            & ---           & ---\\       \hline
Convolution, BN & $16\times(3\times3)$  & 1168, 32      & ReLU\\      \hline
MaxPooling      & $2\times2$            & ---           & ---\\       \hline
Convolution, BN & $32\times(3\times3)$  & 4640, 64      & ReLU\\      \hline
FullyConnected  & 20                    & 2304020       & SoftMax\\   \hline
\end{tabular}
\end{table}

\section{Security Enhancement of RFFI}\label{sec:securityanalysis}
The study of attack and defense of RFFI is relatively limited since RFFI is still in its early stage~\cite{zhang2023radio}. The attacks are varied and can compromise RFFI by attacking the classifiers~\cite{ma2023white}, replicating RFFs~\cite{karunaratne2021penetrating}, and spoofing~\cite{liu2023robust}. It is impractical to analyze all the attacks. Impersonation attacks are analyzed specifically in the context as they are easy to carry since RFFs are constantly exposed and analyzed by adversaries due to the broadcast nature of wireless communications~\cite{lu2024erasing}, and it has been reported that RFFs can be easily impersonated if the discrimination of an RFFI classifier is not well-established~\cite{rehman2014analysis,robyns2017physical}. More importantly, the transfer learning implementation in the proposed RFFI system has a security flaw in the enrollment step, potentially making false RFFs present in the RFF collection.

Implementing the PA nonlinearity quotient and transfer learning classifier involves base model training, enrollment, and authentication. Base model training requires sample collection in a channel effect-free environment, typically during device manufacturing. Therefore, no rogue devices can send false identity information and develop impersonation attacks at this step. Enrollment and authentication steps require sample collection in uncontrolled deployment environments, making them vulnerable to impersonation attacks. As a result, we performed impersonation attacks during the two steps to conduct a security analysis on the PA nonlinearity quotient and transfer learning classifier.

\subsection{Threat Model}\label{sec:threatmodel}
\begin{figure*}[t]
\centering
\includegraphics[width=0.85\textwidth]{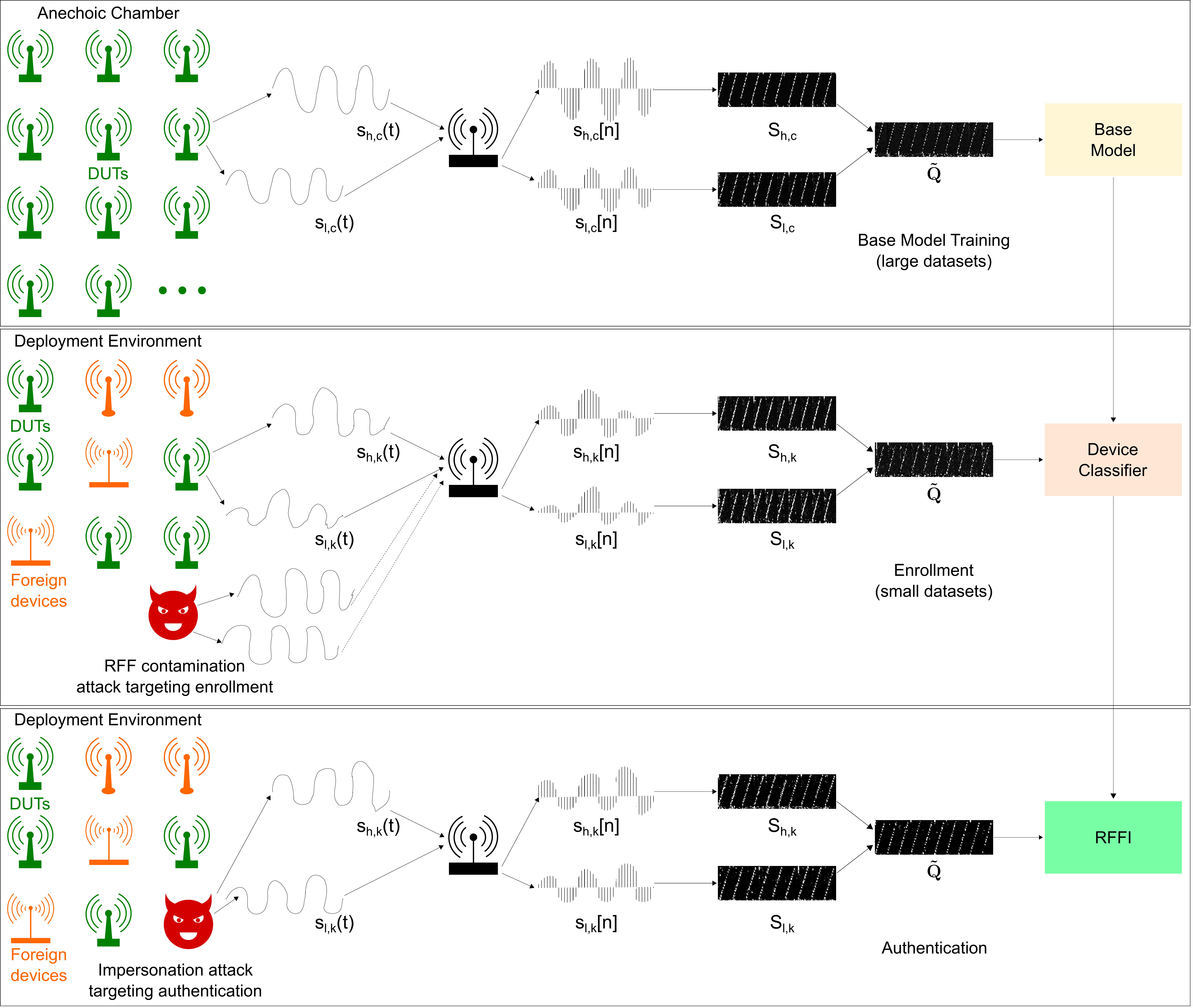}
\caption{The RFF contamination attack in the enrollment step and the impersonation attack in the authentication step.}
\label{fig:attack model}
\end{figure*}

We have identified and formalized two impersonation attacks in the enrollment and authentication steps, as shown in Fig.~\ref{fig:attack model}. The attack during the enrollment process has been labeled as the RFF contamination attack, while the attack during the authentication step has been named the impersonation attack for differentiation.

We adopt the same threat model in~\cite{rehman2014analysis}. In the RFF contamination attack, rogue devices send false identities to contaminate the RFF datasets used for base model retraining. In the impersonation attack, rogue devices send false identities to pass themselves off as legitimate devices. These rogue devices possess knowledge of transmission protocols, message formats, and software addresses necessary to ensure that intended recipients receive false identity messages. Furthermore, the rogue devices can monitor all communications, disrupt legitimate devices, and prevent triggering alerts when the recipients receive duplicate enrollment or authentication messages.

\subsection{Evaluation Metric}\label{sec:ROCcurve}
The security analysis of the PA nonlinearity quotient and transfer learning classifier under the identified attacks was performed using the receiver operating characteristic (ROC) curve. For each class in the classifier, threshold values ranging from 0 to 1 were applied to compare the output probabilities from the fully connected layer. This process calculates each threshold's true-positive rate (TPR) and false-positive rate (FPR). For example, lowering the threshold value classifies more tested samples as positive, thus increasing both TPR and FPR. The ROC curve is drawn upon the TPR and FPR pairs. The area under the ROC curve (AUC) is calculated by integrating the ROC curve with respect to the FPR. AUC values fall from 0 to 1, with a higher value indicating the proposed classifier's more excellent rogue device detection ability and, thus, better security to defend against the identified attacks. Each classifier class has its own AUC value representing the rogue device detection ability of the associated DUT.

\section{Experiments}\label{sec:experiments}

\begin{table}[t]
  \footnotesize
  \centering
  \caption{DUT Configurations}
    \begin{tabular}{|c|c|c|c|c|}
    \hline
    \begin{tabular}[c]{@{}c@{}}Carrier\\Frequency\end{tabular} & \begin{tabular}[c]{@{}c@{}}Bandwidth\\($B$)\end{tabular} & \begin{tabular}[c]{@{}c@{}}Transmission\\Power (h/l)\end{tabular} & \begin{tabular}[c]{@{}c@{}}Spreading\\Factor ($SF$)\end{tabular} & \begin{tabular}[c]{@{}c@{}}Coding\\Rate\end{tabular} \bigstrut\\
    \hline
    915~MHz & 62.5~kHz & 17/10~dBm & 10 & 4/5 \bigstrut\\
    \hline
    \end{tabular}
  \label{tab:config}
\end{table}

\begin{figure}[t]
\centering
\includegraphics[width=0.448\textwidth]{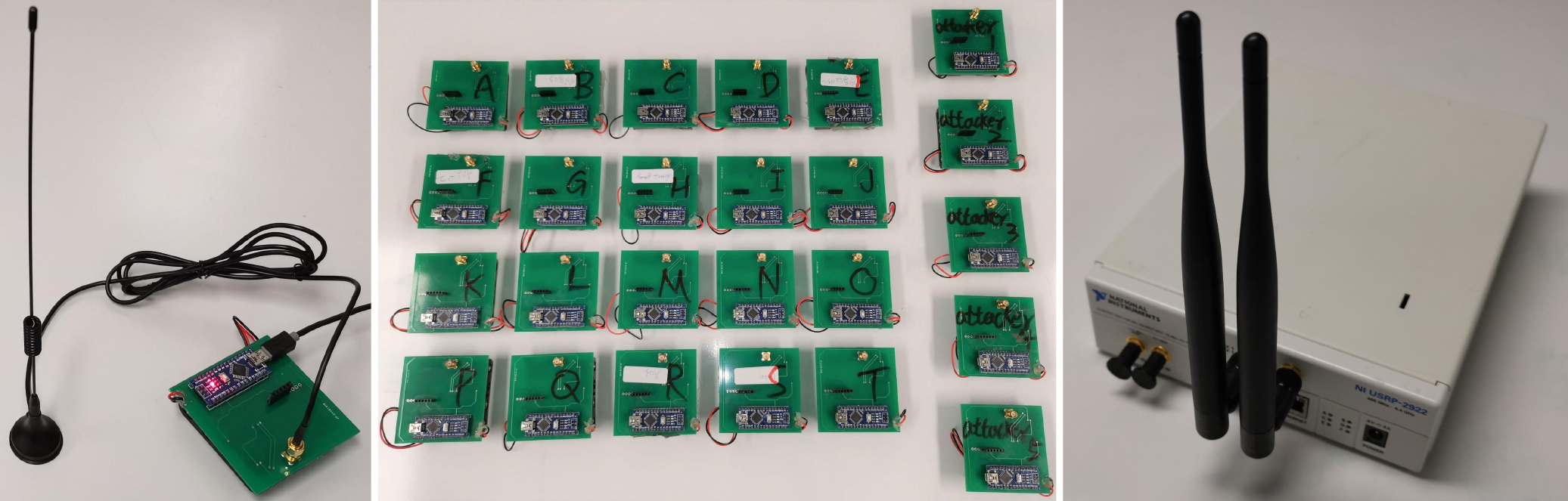}
\caption{Devices employed in the experiments. Left: an operating DUT. Middle: 20 DUTs as legitimate devices and 5 DUTs as rogue devices. Right: a USRP-2922 platform for RFF sample collection.}
\label{fig:devices}
\end{figure}

\begin{figure}[t]
\centering
\includegraphics[width=0.45\textwidth]{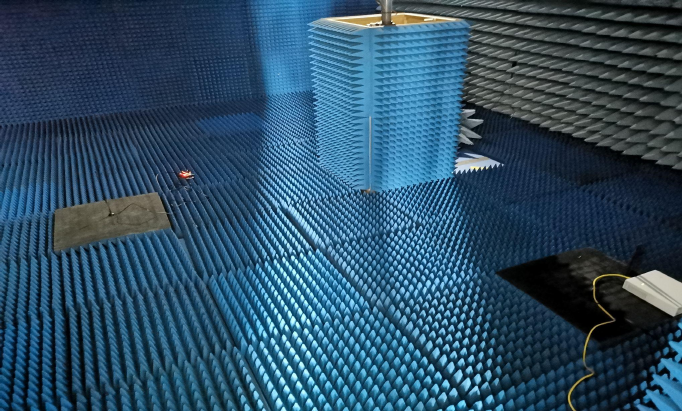}
\caption{Anechoic chamber.}
\label{fig:anechoicCh}
\end{figure}

\begin{figure}[t]
\centering
\includegraphics[width=0.45\textwidth]{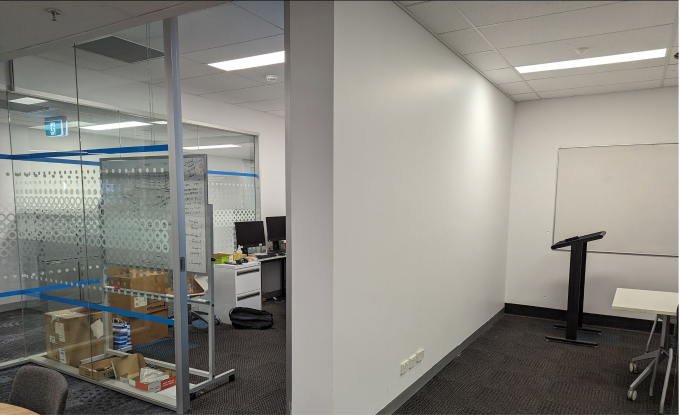}
\caption{Indoor environment.}
\label{fig:indoorEn}
\end{figure}

\begin{figure}[t]
\centering
\includegraphics[width=0.45\textwidth]{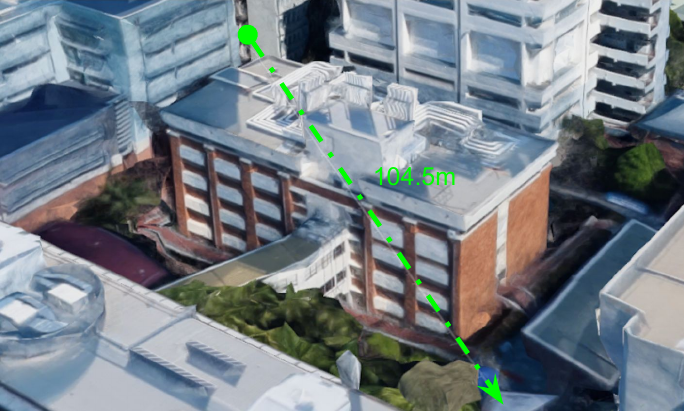}
\caption{Outdoor environment.}
\label{fig:outdoorEn}
\end{figure}

We employed 25 Arduino Nano-controlled LoRa SX1276 modules as the Devices Under Test (DUTs). All DUTs share identical circuit designs and specifications. From these, 20 DUTs were randomly selected and designated as legitimate devices, labeled ``DUT: A" to ``DUT: T," while 5 DUTs were chosen as rogue devices, labeled ``DUT: Attacker 1" to ``DUT: Attacker 5." LoRa offers bandwidths ranging from 7.8~kHz to 500~kHz. However, the minimal bandwidth in practice is 62.5~kHz as bandwidth below the value requires a temperature-compensated crystal oscillator~\cite{jovalekic2018lora}. It is also important to note that the commonly selected 125~kHz, 250~kHz, and 500~kHz bandwidths are caused by the implementation of the legacy LoRaWAN protocol, which should not be confused with the LoRa technology. Other bandwidths, such as 62.5~kHz, are also used in LoRa research~\cite{hu2020sclora,xia2021litenap}. The experiments adopted a bandwidth of 62.5~kHz as it can mitigate packet loss in outdoor transmission. Detailed device configurations are outlined in Table~\ref{tab:config}. To capture RFF samples, we employed a universal software radio peripheral (USRP) platform operating at a sampling frequency of 1~MS/s ($f_S$). The employed devices are shown in Fig.~\ref{fig:devices}. RFF samples were collected in three environments as follows.
\begin{itemize}
    \item \textbf{Anechoic chamber:} channel effect-free RFF samples were collected in the anechoic chamber for training a base model as shown in Fig.~\ref{fig:anechoicCh}. DUTs were placed 3~m away from the USRP platform. The anechoic chamber was designed to absorb multipath signals. Therefore, RFF samples collected in the environment can train the base model that later is exploited to learn channel effects in transfer learning.
    \item \textbf{Indoor environment:} the indoor environment is considered to have moderate multipath fading. DUTs were placed in an office room, and the USRP platform was placed in the adjacent room with a wall in between, as shown in Fig.~\ref{fig:indoorEn}. People were freely walking in the environment in the RFF sample collection.
    \item \textbf{Outdoor environment:} the outdoor environment is considered to have more significant multipath fading than the indoor environment. DUTs were placed 104.5~m from the USRP platform, as shown in Fig.~\ref{fig:outdoorEn}. Buildings blocked the line of sight, and people freely walked in the environment in the RFF sample collection.
\end{itemize}

\begin{figure*}[t]
\centering
\includegraphics[width=0.7\textwidth]{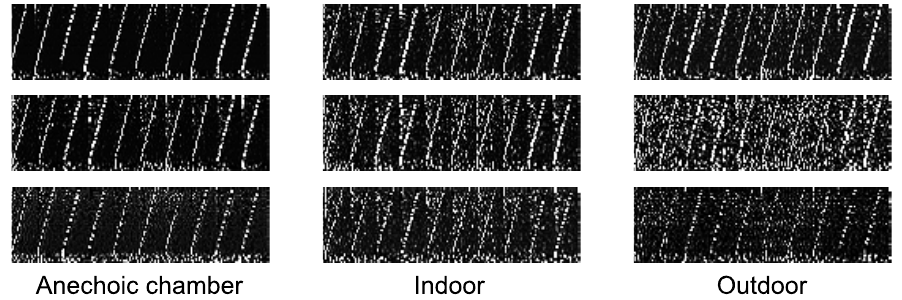}
\caption{The preamble spectrogram and PA nonlinearity quotient of a DUT generated in the anechoic chamber, indoor, and outdoor environments. Top: high-power preamble spectrogram. Middle: low-power preamble spectrogram. Bottom: PA nonlinearity quotient, $\widetilde{\boldsymbol{Q}}$ in~\eqref{eq:QdB}. All in dB scale.}
\label{fig:fingerprintIMG}
\end{figure*}

In the RFF sample collection, DUTs transmitted packets with alternating high-power and low-power modes, and the USRP platform passively received the packets. Over 2800 packets were collected for a DUT in one hour in each environment. Hence, more than 8400 packets were collected for one DUT in three environments. Fig.~\ref{fig:fingerprintIMG} shows the LoRa preamble spectrogram and the PA nonlinearity quotient generated in the anechoic chamber, indoor, and outdoor environments.

\section{Robustness Enhancement Results}\label{sec:classificationanalysis}

\begin{figure}[t]
\centering
\includegraphics[width=0.44\textwidth]{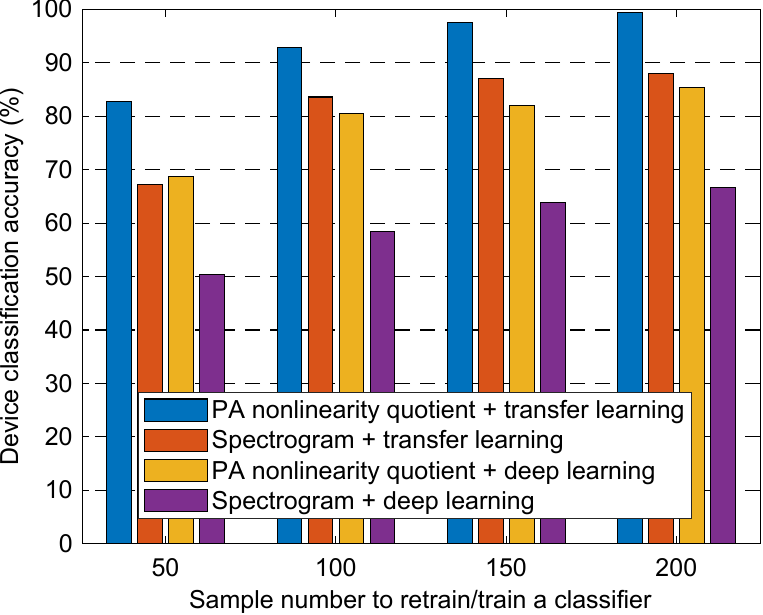}
\caption{Device classification in the indoor environment.}
\label{fig:indoorClassification}
\bigbreak
\includegraphics[width=0.44\textwidth]{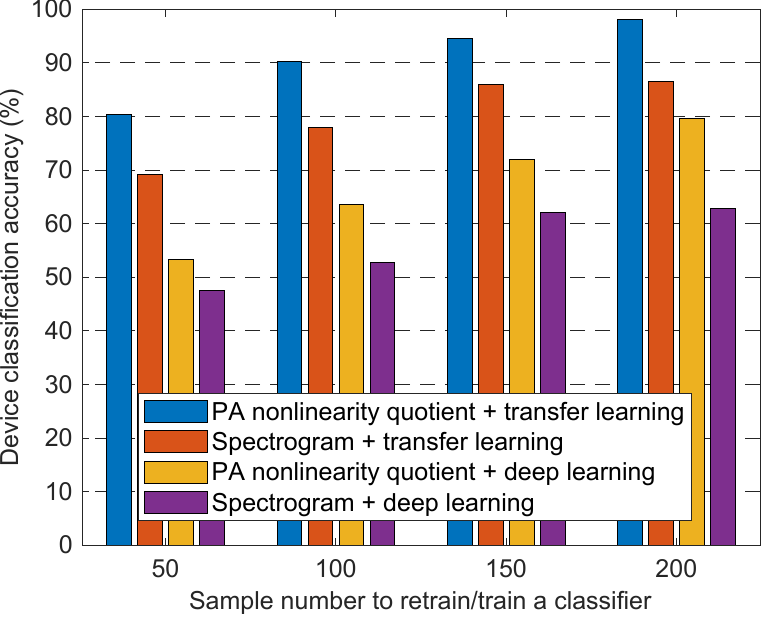}
\caption{Device classification in the outdoor environment.}
\label{fig:outdoorClassification}
\end{figure}
Upon the data collection in the anechoic chamber, indoor, and outdoor environments, we trained transfer learning and deep learning classifiers exploiting the PA nonlinearity quotient and preamble spectrogram to validate the significance of the proposed classifier in device classification. A base classifier underwent training using complete datasets from legitimate devices (DUTs: ``A" to ``T") collected in the anechoic chamber. Subsequently, small training sets, with sample counts of 50, 100, 150, and 200,  were generated from the indoor and outdoor environments for transfer learning. Spectrogram and deep learning classifiers were trained with spectrogram samples as the benchmark. We utilized identical test sets from indoor and outdoor environments to evaluate the effectiveness of the proposed PA nonlinearity quotient and transfer learning classifier, along with spectrogram and deep learning classifiers, each containing over 1000 samples per DUT. These test sets were separate from the training sets and used for validation.

Fig.~\ref{fig:indoorClassification} illustrates device classification results in the indoor environment. The proposed PA nonlinearity quotient and transfer learning classifier outperformed the spectrogram and deep learning classifier, showcasing an average classification accuracy enhancement of $33.3\%$. More samples for retraining the base model leads to higher classification accuracy. An excellent $99.4\%$ accuracy was achieved by retraining the base classifier with only 200 samples. Further, the PA nonlinearity quotient yielded an average classification accuracy improvement of $19.4\%$ compared to the preamble spectrogram. Fig.~\ref{fig:outdoorClassification} depicts device classification results in the outdoor environment. Like the indoor scenario, the proposed PA nonlinearity quotient and transfer learning classifier outperformed the spectrogram and deep learning classifier. The average classification accuracy improvement is $34.5\%$. Further, the PA nonlinearity quotient in the outdoor environment contributed to an average classification accuracy improvement of $10.9\%$ compared to the preamble spectrogram.

Table~\ref{table:DeviceClassificationComparison} compares device classification performance among the proposed PA nonlinearity quotient and transfer learning classifier and notable works in literature. The proposed classifier achieved high accuracy in classifying devices in indoor and outdoor environments, using a few training samples and requiring less power and memory storage for RFFI.
\begin{table}[t]
\footnotesize
\centering
\caption{Device Classification Comparison}
\label{table:DeviceClassificationComparison}
\begin{tabular}{|c|c|c|c|c|}
\hline
Work & \begin{tabular}[c]{@{}c@{}}Experimental\\Environments\end{tabular} & \begin{tabular}[c]{@{}c@{}}No. of\\Devices\end{tabular} &\begin{tabular}[c]{@{}c@{}}Training Samples\\(Per Device)\end{tabular} & \begin{tabular}[c]{@{}c@{}}Classification\\Accuracy\end{tabular} \\\hline
Ours                  & \begin{tabular}[c]{@{}c@{}}Indoor\\Outdoor\end{tabular} & 20 & 200 & \begin{tabular}[c]{@{}c@{}}99.4\%\\98.2\%\end{tabular} \\\hline
\cite{shen2022towards}   & Indoor   & 30    & 100   & 98.4\%    \\\hline
\cite{yu2019robust}      & Indoor   & 54    & 698   & 84.6\%    \\\hline
\cite{xing2022design}    & Indoor   & 7     & 800   & 99.0\%    \\\hline
\end{tabular}
\end{table}

\section{RFFI Security Analysis Results}\label{sec:impersonationattackanalysis}
\subsection{Impersonation Attack Targeting Authentication}

\begin{figure}[t]
\centering
\includegraphics[width=0.4\textwidth]{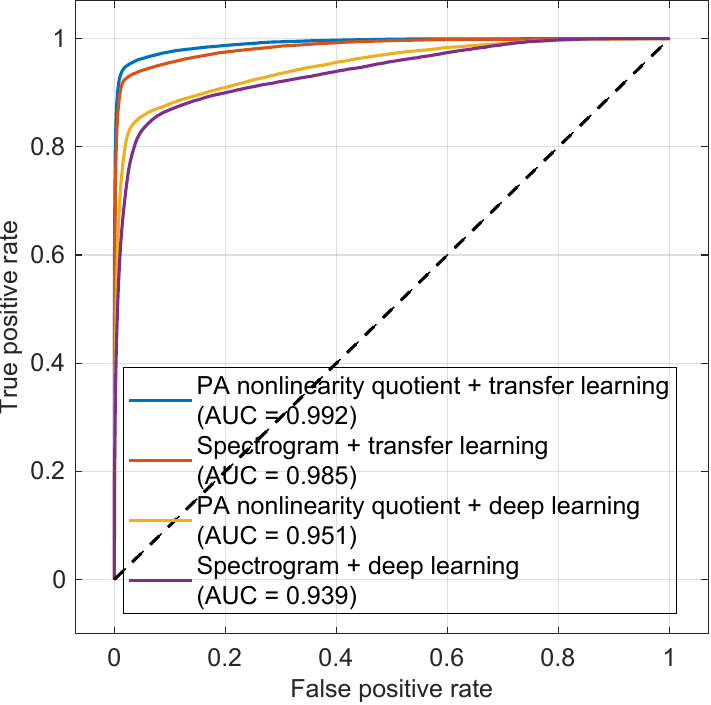}
\caption{ROC curves for the impersonation attack detection in the indoor environment.}
\label{fig:indoorROC}
\bigbreak
\includegraphics[width=0.4\textwidth]{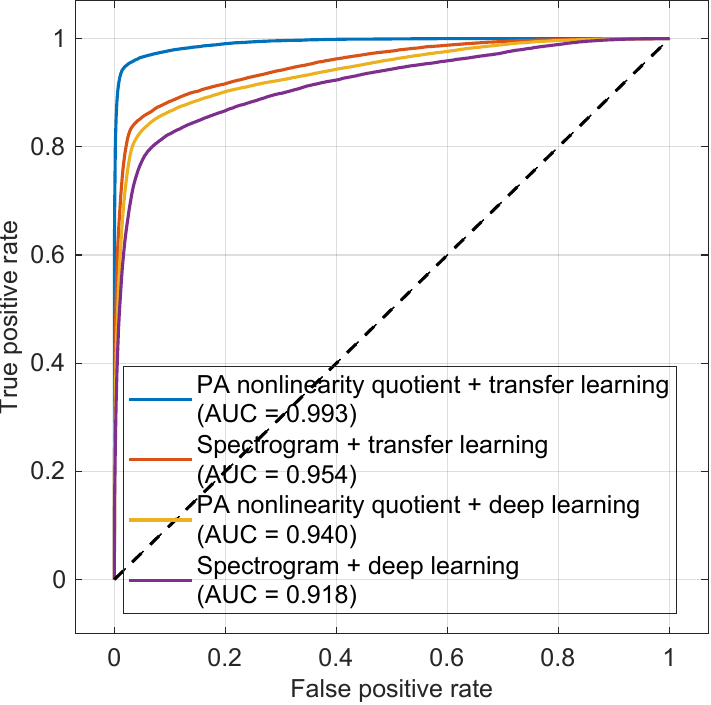}
\caption{ROC curves for the impersonation attack detection in the outdoor environment.}
\label{fig:outdoorROC}
\end{figure}

To study the impact of the impersonation attack in the authentication step, we randomly selected 100 RFF samples per legitimate device (DUT: ``A'' to ``T'') to build a dataset for retraining the base model in transfer learning. Spectrogram and deep learning classifiers were also trained with the same dataset as the benchmark. A test set consisted of over 1000 RFF samples of a legitimate device under the attack and more than 1000 for each rogue device (DUT: ``Attacker~1'' to ``Attacker~5''), totaling 6000 samples. The rogue devices claimed identity as the targeted DUT to develop the impersonation attack. The attack was developed among all legitimate devices, and a micro-averaging method, which treats all one-versus-all binary classification problems as one binary classification problem, was used to generate average ROC and AUC values. None of the samples from the training set were included in the test set.

The ROC curves for the impersonation attack detection in the indoor environment are presented in Fig.~\ref{fig:indoorROC}. The PA nonlinearity quotient and transfer learning classifier outperformed the spectrogram and deep learning classifier, achieving an AUC value of 0.992 compared to 0.939. Moving on to the outdoor environment, as shown in Fig.~\ref{fig:outdoorROC}, the proposed classifier demonstrated excellent performance in rogue device detection, achieving an AUC value of 0.993 compared to 0.918. Further, the PA nonlinearity quotient exhibited greater robustness to environmental changes than the spectrogram, yielding larger AUC values and better rogue device detection capability in indoor and outdoor experiments. Overall, the PA nonlinearity quotient and transfer learning classifier demonstrated excellent rogue device detection capability for the impersonation attack and significantly outperformed the conventional spectrogram and deep learning classifier.

\subsection{RFF Contamination Attack Targeting Enrollment}\label{sec:RFFcontaminationattackanalysis}

\begin{figure}[t]
\centering
\includegraphics[width=0.4\textwidth]{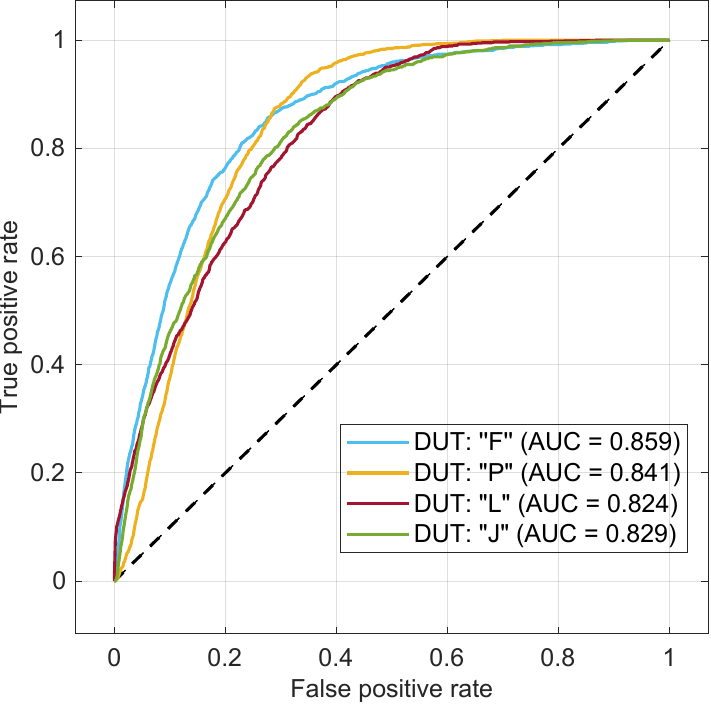}
\caption{ROC curves for the RFF contamination attack detection in the indoor environment.}
\label{fig:indoor RFF contamination attack}
\bigbreak
\includegraphics[width=0.4\textwidth]{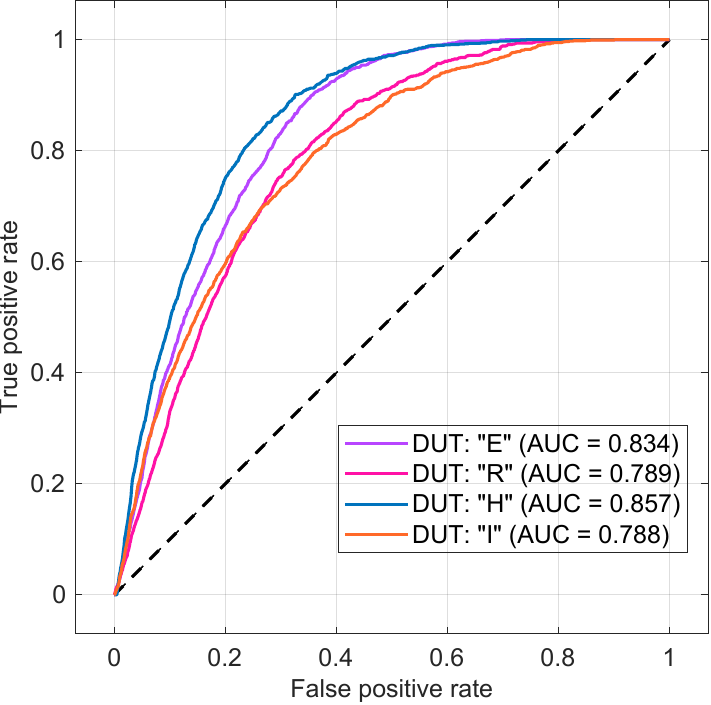}
\caption{ROC curves for the RFF contamination attack detection in the outdoor environment.}
\label{fig:outdoor RFF contamination attack}
\end{figure}

The RFF contamination attack is practical because the RFF samples in the enrollment step for retraining the base model must be collected in uncontrolled deployment environments to acknowledge real channel effects and enhance RFFI robustness. The threat model allows receivers to receive false identity messages without being detected through disruption.

To study the impact of the RFF contamination attack in the enrollment step, we randomly selected 100 RFF samples per legitimate device (DUT: ``A'' to ``T'') to build a dataset for retraining the base model in transfer learning. The RFF contamination attack was carried out by randomly selecting one legitimate device as the attack target, with the RFF samples for retraining the base model replaced by RFF samples from an arbitrary rogue device (DUT: ``Attacker~1'' to ``Attacker~5''). A test set consisted of over 1000 samples from the targeted device and over 1000 from the rogue device. None of the samples from the training set were in the test set.

The experiment included 20 legitimate devices and 5 rogue devices, so there were 100 attacking scenarios. It would be labor-intensive to demonstrate the impact of the RFF contamination attack on all 20 valid devices. As a result, we randomly chose four valid devices from each environment for demonstration purposes. It is important to note that all the devices have the same circuit design and configurations. The impact of the RFF contamination attack on the four devices can be applied to the other devices.

DUT ``F'', ``P'', ``L'', and ``J'' were randomly selected as the attack targets in the indoor environment. DUT ``E'', ``R'', ``H'', and ``I'' were randomly selected as the attack targets in the outdoor environment. The ROC curves for the RFF contamination attack detection in the indoor environment are presented in Fig.~\ref{fig:indoor RFF contamination attack}. The PA nonlinearity quotient and transfer learning classifier is vulnerable to the RFF contamination attack since all AUC values are small, and AUC values below 0.9 are considered to have no outstanding discrimination ability in statistics~\cite{bradley1997use}. A similar observation applied to the outdoor environment as shown in Fig.~\ref{fig:outdoor RFF contamination attack}. Therefore, a countermeasure is required to defend against the RFF contamination attack in the enrollment step for a secure RFFI application.

\section{Lightweight Keyless Countermeasure to RFF Contamination Attack}\label{sec:attackcountermeasure}

The PA nonlinearity quotient and transfer learning classifier was vulnerable to the RFF contamination attack in the experiments. It should be noted that this vulnerability is present in other RFFI techniques that require data collection in uncontrolled environments. Cryptography-based authentication techniques are used to defend against attacks carried out with false identities~\cite{ferrag2017authentication}. However, the use of public key cryptography and preshared keys compromise the non-cryptographic nature of RFFI. Therefore, we propose a keyless countermeasure to defend against the attacks targeting RFFI by sending RFF with false identities. The countermeasure exploits probability information automatically obtained from model training, with no extra overheads, and thus can be readily applied to IoT applications.

\subsection{Attack Detection with Posterior Classification Probability}\label{sec:probabilitydifference}

\begin{figure*}[t]
\centering
\includegraphics[width=0.95\textwidth]{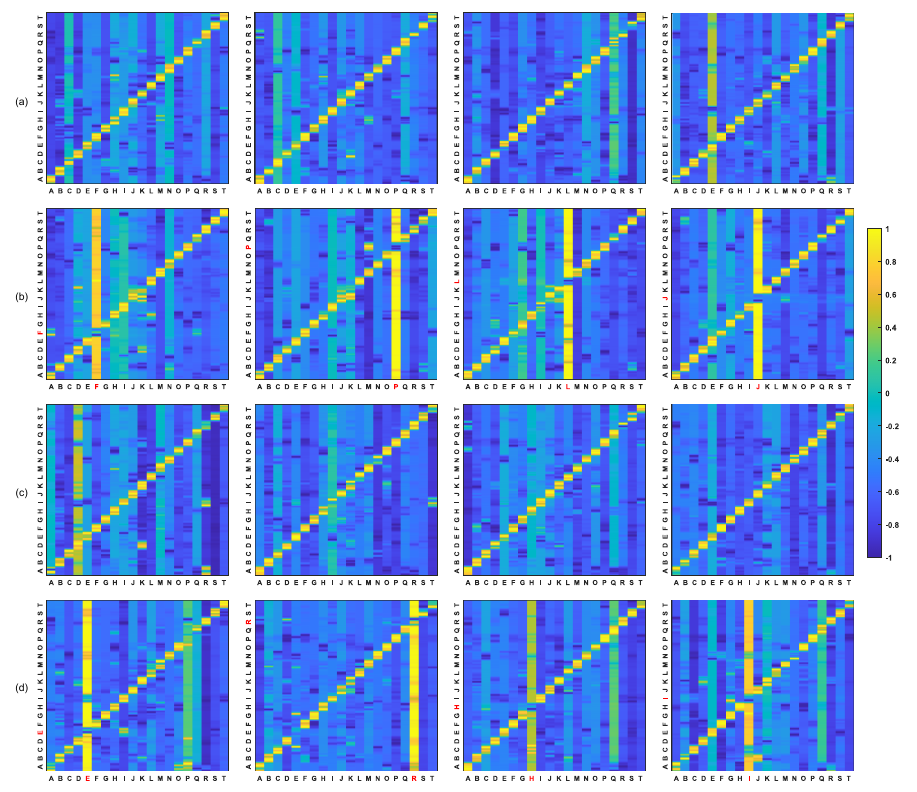}
\caption{$\hat{\mathcal{M}}_{diff}$ matrices in heat maps. Classifiers retrained/trained with 100 RFF samples per DUT. (a). Indoor environment with no rogue devices. (b). Indoor environment with rogue devices targeting DUT: ``F'', ``P'', ``L'', and ``J'', respectively. (c). Outdoor environment with no rogue devices. (d). Outdoor environment with rogue devices targeting DUT: ``E'', ``R'', ``H'', and ``I'', respectively.}
\label{fig:Pdifference}
\end{figure*}

In machine learning, multilayer perceptrons are naturally probabilistic. The output of the softmax function is a matrix of posterior probabilities ($\mathcal{M}$), with the number of rows equal to the number of observations ($O$) and the number of columns equal to the number of classes ($C$), where $o=1:O$ is used to index observations and $c=1:C$ is used to index classes. The element $(o,c)$ is the posterior probability that observation $o$ belongs to class $c$, which is expressed as $\Pr(c\mid o)$.

To implement the keyless countermeasure, two classifiers need to be trained using RFF samples collected in uncontrolled deployment environments for enrollment. One classifier uses transfer learning and is fine-tuned on a pre-trained model obtained from the base model training step. The other classifier uses deep learning and is trained from scratch. The training of the classifiers takes place at the receivers and adds no overhead to the tested devices.

It is demonstrated in Section~\ref{sec:classificationanalysis} that the pre-trained base model had the knowledge of RFFs from legitimate devices that improved the device classification in the transfer learning classifier in contrast with the deep learning classifier trained from scratch. Therefore, when no RFF contamination attack happens, the posterior probabilities at true-positive pairs of observations and classes resulting from the transfer learning classifier are higher than the deep learning classifier.

With $(\mathbf{o},\mathbf{c})$ are pairs of RFF samples and corresponding devices and no rogue devices in the enrollment, the test set can be expressed as $\mathcal{D}=\{(\mathbf{o}^{(i)},\mathbf{c}^{(i)})\mid i=1:N\}$, where $N$ is the number of testing pairs. The posterior probability difference produced between the transfer learning and deep learning classifier in the experiments from $\mathcal{D}$ is given as
\begin{align*}
\Pr_{transfer}(\mathbf{c}^{(i)}\mid \mathbf{o}^{(i)})-\Pr_{deep}(\mathbf{c}^{(i)}\mid \mathbf{o}^{(i)})>0,
\label{eq:probabilityDiffereNo}
\tag{8}
\end{align*}
where $\underset{transfer}{\Pr}(\cdot)$ and $\underset{deep}{\Pr}(\cdot)$ are probability operators for the transfer learning and deep learning classifiers, respectively.

The relation established in~\eqref{eq:probabilityDiffereNo} changes as the RFF contamination attack happens, making an indication in attack detection. Transfer learning tunes weights and bias of a pre-trained base model slowly with a small learning rate. Since RFF samples are distinctive among legitimate and rogue devices, tuning the base model with a few RFF samples from rogue devices results in mixed feature representations. Thus, the resulting classifier has limited capability to classify legitimate or rogue devices. In contrast, deep learning has a large learning rate that changes weights and biases faster than transfer learning. The resulting classifier has no mixed feature representations when training with the same RFF samples from rogue devices. It can classify rogue devices more accurately than legitimate devices. Therefore, the posterior probability at true-positive pairs of the legitimate device under attack from a transfer learning classifier is lower than a deep learning classifier. For the test set $\breve{\mathcal{D}}=\{(\mathbf{\breve{o}}^{(i)},\mathbf{\breve{c}}^{(i)})\mid i=1:N\}$, where $(\mathbf{\breve{o}},\mathbf{\breve{c}})$ are the observations and corresponding devices targeted by the RFF contamination attack, the posterior probability difference is given as
\begin{align*}
\Pr_{transfer}(\mathbf{\breve{c}}^{(i)}\mid \mathbf{\breve{o}}^{(i)})-\Pr_{deep}(\mathbf{\breve{c}}^{(i)}\mid \mathbf{\breve{o}}^{(i)})<0.
\label{eq:probabilityDiffereYes}
\tag{9}
\end{align*}

Since the softmax function automatically generates posterior probability matrices, we calculate the posterior probability difference by subtracting the softmax output of a deep learning classifier from a transfer learning classifier. The matrix of the posterior probability difference is given as
\begin{align*}
\mathcal{M}_{diff} = \mathcal{M}_{Transfer} - \mathcal{M}_{Deep},
\label{eq:PrDifference}
\tag{10}
\end{align*}
where $\mathcal{M}_{Transfer}$ and $\mathcal{M}_{Deep}$ are test set outputs of the softmax function from a transfer learning classifier and a deep learning classifier, respectively. $\mathcal{M}_{diff}$ is normalized in $[-1,1]$, which is denoted as $\hat{\mathcal{M}}_{diff}$.

\subsection{Attack Detection Analysis}\label{sec:experimentalanalysisdifference}
The use of $\hat{\mathcal{M}}_{diff}$ occurs during the enrollment step to construct an RFFI system using the PA nonlinearity quotient and transfer learning classifier. All the necessary processes to generate $\hat{\mathcal{M}}_{diff}$ take place at the receivers, adding no extra overhead to DUTs.

Algorithm~\ref{algorithm:RFFContaminationAttackDetection} demonstrates the generation of $\hat{\mathcal{M}}_{diff}$ at the receivers. RFFs collected from DUTs in deployment environments for enrollment are divided into training and test sets. Two classifiers are then trained using the same training set. This includes a transfer learning classifier, which is trained by fine-tuning a pre-trained base model, and a deep learning classifier trained from scratch. Notably, the pre-trained base model is trained with RFFs collected from the DUTs in controlled environments without channel effects and adversaries.

The test set is then used to generate posterior probability matrices from the softmax function. $\hat{\mathcal{M}}_{diff}$ is derived by computing the difference in posterior probability between the transfer learning and deep learning classifiers. If no attacks are detected, the transfer learning classifier used in the attack detection can be further utilized for RFFI.

\begin{algorithm}[t]
\footnotesize
\DontPrintSemicolon
  \KwInput{$\mathcal{T},\mathcal{D}$ \quad $\%$Training and testing RFF sample sets}
  \KwInput{$\mathcal{H}_B$ \quad $\%$Base model}
  \KwOutput{$\hat{\mathcal{M}}_{diff}$ \quad $\%$Normalized posterior probability difference}
  Transfer learning: $\mathcal{H}_B,\mathcal{T}\rightarrow \mathcal{H}_T$ \quad $\%$Transfer learning classifier\\
  Deep learning: $\mathcal{T}\rightarrow \mathcal{H}_D$ \quad $\%$Deep learning classifier\\
  $\mathcal{H}_T,\mathcal{D}\xrightarrow{softmax}\mathcal{M}_{T}$ \quad $\%$Posterior probability matrix\\
  $\mathcal{H}_D,\mathcal{D}\xrightarrow{softmax}\mathcal{M}_{D}$\\
  $\hat{\mathcal{M}_{diff}}=normalize(\mathcal{M}_{T} - \mathcal{M}_{D})$\\
\caption{Posterior Probability Difference.}
\label{algorithm:RFFContaminationAttackDetection}
\end{algorithm}

To validate $\hat{\mathcal{M}}_{diff}$ for the RFF contamination attack detection, we developed experiments with 100 RFF samples per DUT for classifier training and 100 pairs of observations and classes per DUT for testing. No training set samples were in the test set. Since there was too much effort to demonstrate the RFF contamination attack detection with all the legitimate devices, the same attack targets in Section~\ref{sec:RFFcontaminationattackanalysis} were selected, including DUT ``F'', ``P'', ``L'', and ``J'' in the indoor environment and DUT ``E'', ``R'', ``H'', and ``I'' in the outdoor environment. The attack detection study was conducted with the worst-case scenario of one legitimate device under the RFF contamination attack each time since it is small-scale and harder to detect. Fig.~\ref{fig:Pdifference} shows $\hat{\mathcal{M}}_{diff}$ matrices using heat maps. It can be seen that the probability difference at true-positive pairs is larger than zero when no attacks happen, which is consistent with~\eqref{eq:probabilityDiffereNo}. When the RFF contamination attack happens, the probability difference is significantly below zero at true-positive pairs of targeted devices, consistent with~\eqref{eq:probabilityDiffereYes}. Hence, $\hat{\mathcal{M}}_{diff}$ is exploited to detect the RFF contamination attack.

\begin{figure*}[t]
\centering
\subfloat[]{\includegraphics[width=3in]{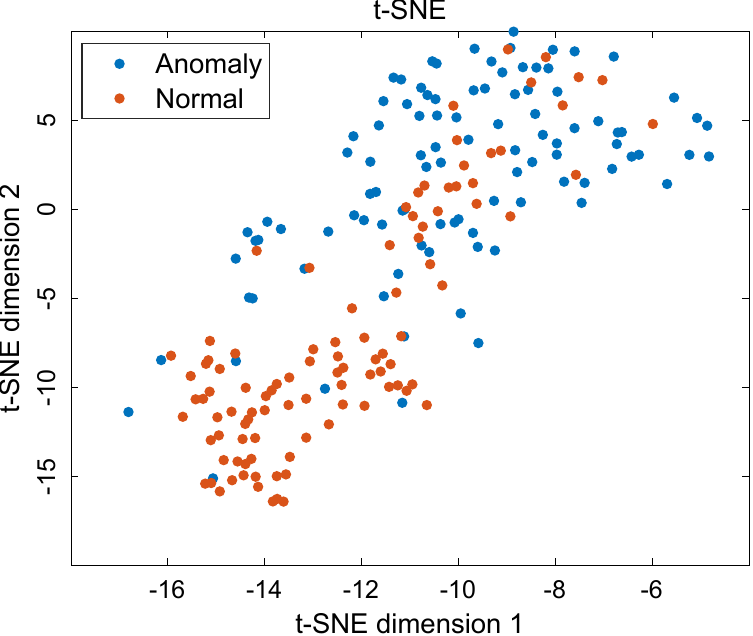}}\hspace{0.04\textwidth}
\subfloat[]{\includegraphics[width=3in]{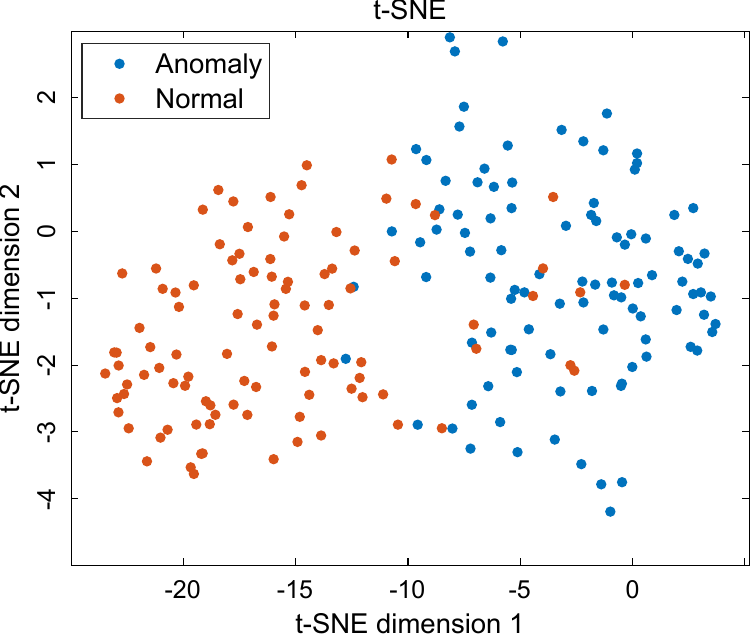}}

\subfloat[]{\includegraphics[width=3in]{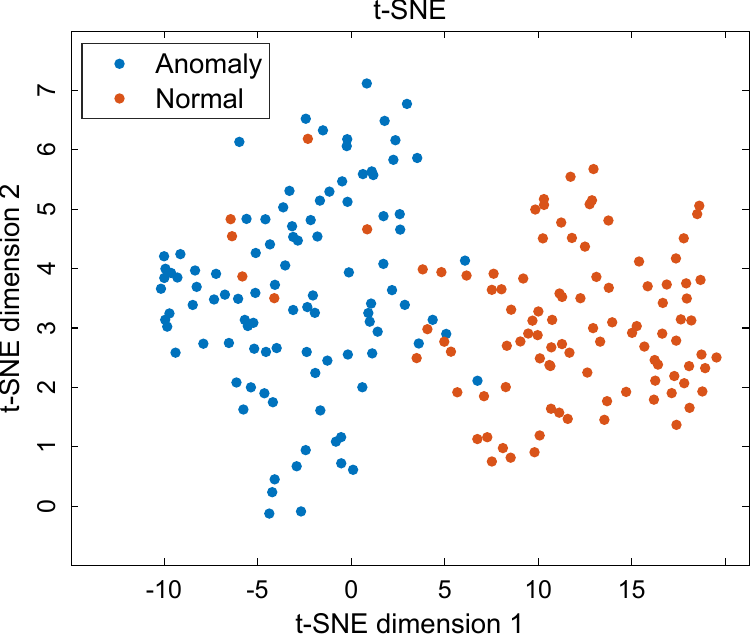}}\hspace{0.04\textwidth}
\subfloat[]{\includegraphics[width=3in]{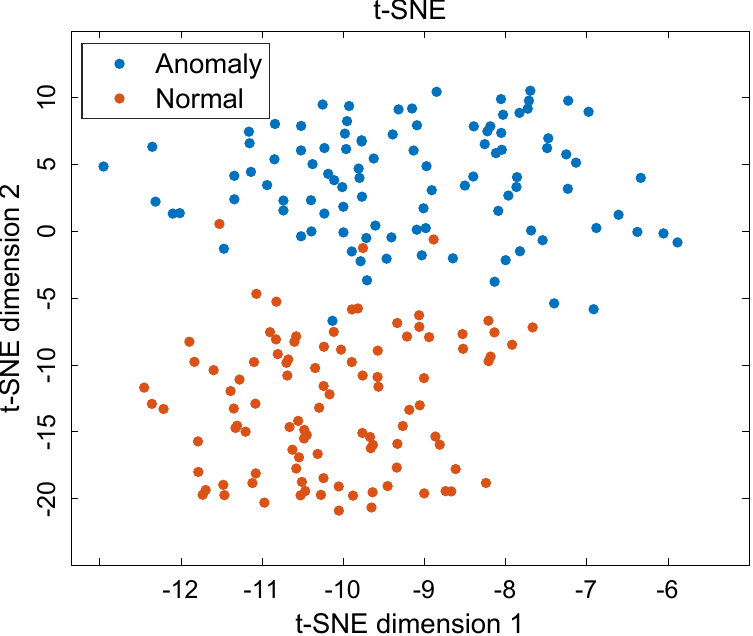}}
\caption{t-SNE visualizations of feature representations of $\hat{\mathcal{M}}_{diff}$ matrices generated in the indoor environment. (a). 50 RFF samples per DUT in classifier training. (b). 100 RFF samples per DUT in classifier training. (c). 150 RFF samples per DUT in classifier training. (d). 200 RFF samples per DUT in classifier training.}
\label{fig:t-SNE_indoor}
\end{figure*}

\begin{figure*}[t]
\centering
\subfloat[]{\includegraphics[width=3in]{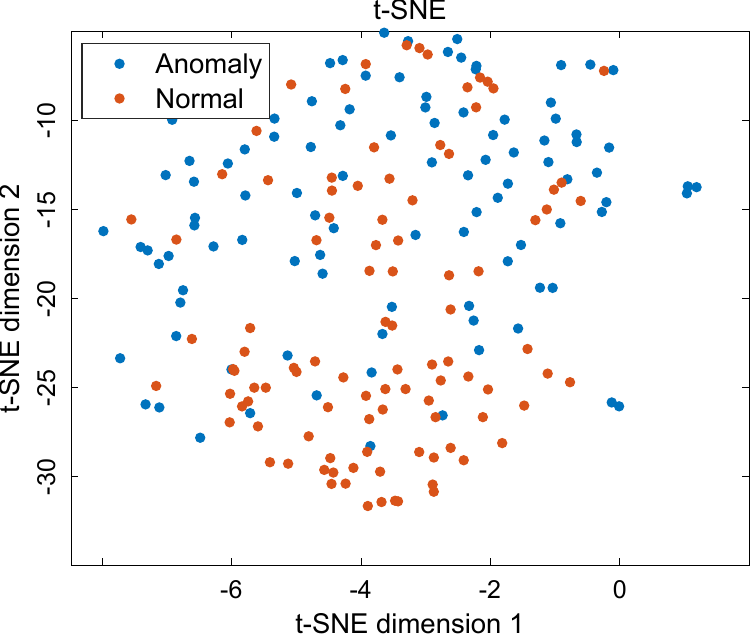}}\hspace{0.04\textwidth}
\subfloat[]{\includegraphics[width=3in]{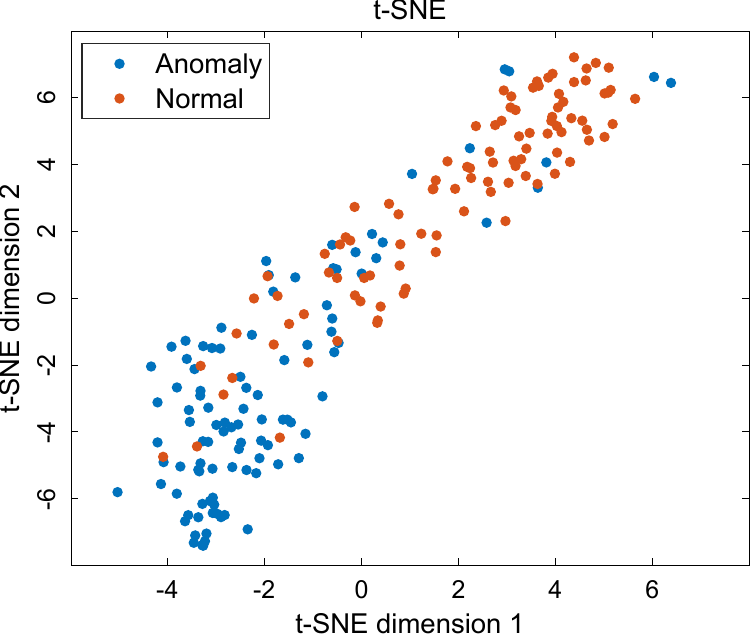}}

\subfloat[]{\includegraphics[width=3in]{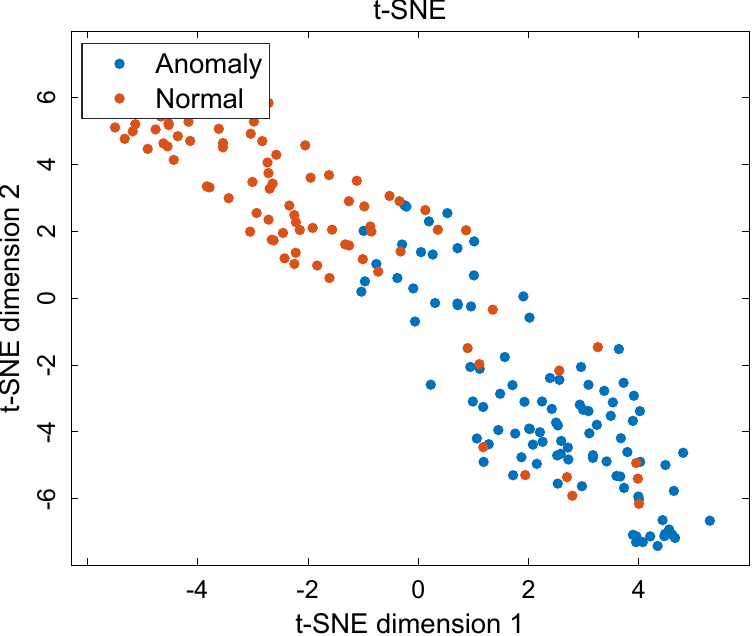}}\hspace{0.04\textwidth}
\subfloat[]{\includegraphics[width=3in]{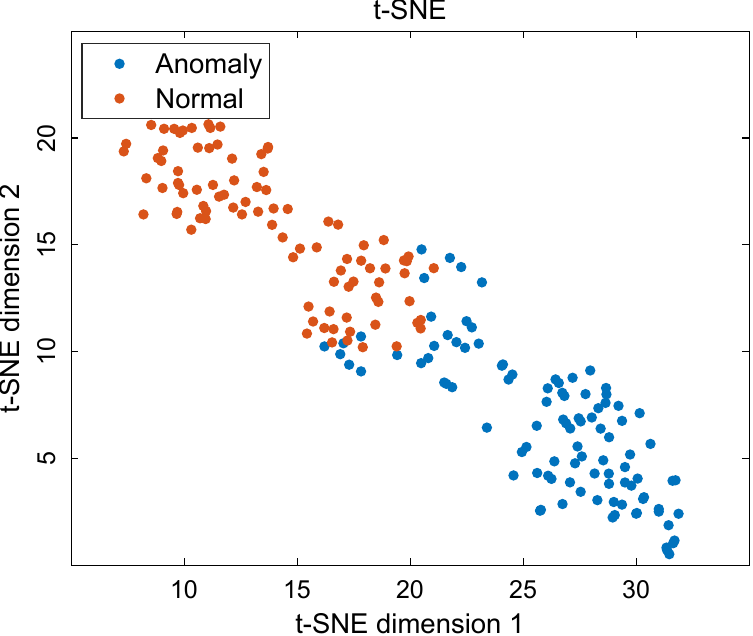}}
\caption{t-SNE visualizations of feature representations of $\hat{\mathcal{M}}_{diff}$ matrices generated in the outdoor environment. (a). 50 RFF samples per DUT in classifier training. (b). 100 RFF samples per DUT in classifier training. (c). 150 RFF samples per DUT in classifier training. (d). 200 RFF samples per DUT in classifier training.}
\label{fig:t-SNE_outdoor}
\end{figure*}
Attack detection can be developed with normal and abnormal case classification. However, generating all abnormal $\hat{\mathcal{M}}_{diff}$ matrices is challenging. We propose to use a one-class support vector machine (SVM) for the attack detection since the no attack case and corresponding $\hat{\mathcal{M}}_{diff}$ matrices are readily exploited. The one-class SVM algorithm can develop outlier detection with training on normal data~\cite{ruff2021unifying}. Specifically, it computes and defines a regime boundary that accommodates most training data. If test samples fall in the regime, they are categorized as normal operating data points. Otherwise, they are outliers and detected anomalies. Implementing the one-class SVM with AlexNet architecture is widely studied in anomaly detection~\cite{oza2019one,perera2019learning,li2023adaptive}. Therefore, we provide the AlexNet architecture with normal data, e.g., Fig.~\ref{fig:Pdifference} (a) and (c), to extract features for a one-class SVM. It is worth noting that many state-of-the-art network architectures can be implemented for better feature extraction than the AlexNet architecture. However, the purpose of the paper is to validate and demonstrate $\hat{\mathcal{M}}_{diff}$ in the RFF contamination attack detection rather than determining the best feature extractors.

We generated 300 normal $\hat{\mathcal{M}}_{diff}$ matrices with no rogue devices in each experimental environment to train the AlexNet. Further, 100 matrices with and 100 matrices without the RFF contamination attack were generated for testing. The RFF contamination attack was developed by deploying a random rogue device (DUT: ``Attacker~1'' to ``Attacker~5'') to attack a random legitimate device (DUT: ``A'' to ``T''). The transfer learning and deep learning classifiers use 50, 100, 150, and 200 RFF samples per DUT for the training.

Fig.~\ref{fig:t-SNE_indoor} shows the visualization of high-dimensional features of $\hat{\mathcal{M}}_{diff}$ matrices generated in the indoor environment using t-distributed stochastic neighbor embedding (t-SNE). Normal matrices represent no attacks, and anomaly matrices represent the RFF contamination attack. The t-SNE is an algorithm to reduce dimensionality. High-dimensional data can be visualized on low-dimensional graphs with the same similarities. It can be observed that the normal and anomaly matrices are distinctive when the RFF samples for training reach 100 per DUT. More samples lead to more distinctive clusters and significant separation boundaries. $\hat{\mathcal{M}}_{diff}$ generated in the outdoor environment has the same observation as in Fig.~\ref{fig:t-SNE_outdoor}.

Table~\ref{table:AUCIndoorDetection} and Table~\ref{table:AUCOutdoorDetection} show the AUC values for Fig.~\ref{fig:t-SNE_indoor} and Fig.~\ref{fig:t-SNE_outdoor} and the corresponding attack detection rate generated with a one-class SVM. An AUC value close to one represents significant discrimination. It can be observed that there is discrimination, with AUC values above 0.98, when RFF samples for retraining reach 200 per DUT. It leads to distinctive clusters of normal and anomaly cases, resulting in more separable observations for the RFF contamination attack. The attack detection rates, defined as the percentage of detected anomaly matrices out of the 100 tested anomaly matrices, reflect the AUC values and show more than $90\%$ when RFF samples for retraining reach 200 per DUT. It is worth noting that adding more RFF samples increases the attack detection rate. However, the energy cost in the enrollment should be considered for power-constrained device applications.

The high-dimensional features visualization, AUC values, and attack detection rates generated with a one-class SVM validated that $\hat{\mathcal{M}}_{diff}$ changes distinctively when rogue devices develop the RFF contamination attack in the enrollment step, resulting an indication for attack detection. Therefore, the keyless defense measure for RFFI systems implemented in uncontrolled environments can be developed exploiting $\hat{\mathcal{M}}_{diff}$ to mitigate the threats of root key compromises and restrictions on secret key renewals.

\begin{table}[t]
  \footnotesize
  \centering
  \caption{Corresponding AUC Values and Attack Detection Rates Generated with a One-Class SVM in Fig.~\ref{fig:t-SNE_indoor}}
    \label{table:AUCIndoorDetection}
    \begin{tabular}{|c|c|c|c|c|}
    \hline
    Samples per DUT & 50     & 100       & 150      & 200   \\ \hline
    AUC             & 0.716  & 0.846     & 0.906    & 0.987 \\ \hline
    Detection Rate  & 63\%   & 75\%      & 84\%     & 93\%  \\ \hline
    \end{tabular}
\end{table}

\begin{table}[t]
  \footnotesize
  \centering
  \caption{Corresponding AUC Values and Attack Detection Rates Generated with a One-Class SVM in Fig.~\ref{fig:t-SNE_outdoor}}
    \label{table:AUCOutdoorDetection}
    \begin{tabular}{|c|c|c|c|c|}
    \hline
    Samples per DUT & 50     & 100      & 150      & 200   \\ \hline
    AUC             & 0.639  & 0.830    & 0.873    & 0.981 \\ \hline
    Detection Rate  & 55\%   & 72\%     & 80\%     & 91\%  \\ \hline
    \end{tabular}
\end{table}

\section{Conclusion}\label{sec:conclusion}
This paper introduces a new channel-robust technique to enhance RFFI performance in changing environments. The experiment results demonstrated significant enhancement in the device classification accuracy with an average classification
accuracy improvement of 33.3\% in indoor environments and
34.5\% in outdoor environments. The paper then addresses the security threat introduced by RFF collection in uncontrolled deployment environments. A security analysis is developed in terms of a formalized impersonation attack. The impersonation attack was found to contaminate RFFs collected for classifier training and ultimately compromised RFFI systems. A novel keyless countermeasure has been proposed to defend against the impersonation attack without sacrificing the lightweight nature of RFFI. The proposed countermeasure has shown excellent attack detection capability with a 40.0\% improvement in attack detection rate in indoor and outdoor environments to enhance the RFFI systems in uncontrolled deployment environments.

\bibliographystyle{IEEEtran}
\bibliography{bib.bib}

\end{document}